\documentclass[a4paper,fleqn]{cas-sc}

\usepackage[numbers]{natbib}
\usepackage{soul}
\usepackage[normalem]{ulem}

\usepackage{bm}
\usepackage{ascmac}

\def\tsc#1{\csdef{#1}{\textsc{\lowercase{#1}}\xspace}}
\tsc{WGM}
\tsc{QE}
\tsc{EP}
\tsc{PMS}
\tsc{BEC}
\tsc{DE}

\begin{document}
\let\WriteBookmarks\relax
\def\floatpagepagefraction{1}
\def\textpagefraction{.001}
\shorttitle{}
\shortauthors{H. Tajiri {\it et~al.}}

\title [mode = title]{Structural insight using anomalous XRD into Mn$_2$CoAl Heusler alloy films grown by magnetron sputtering, IBAS and MBE techniques}                      


\author[1]{Hiroo Tajiri}
\cormark[1]
\ead{tajiri@spring8.or.jp}
\address[1]{Center for Synchrotron Radiation Research, Japan Synchrotron Radiation Research Institute, Hyogo 679-5198, Japan}

\author[1]{Loku Singgappulige Rosantha Kumara}[orcid=0000-0001-9160-6590]

\author[2,3]{Yuya Sakuraba}[orcid=0000-0003-4618-9550]
\address[2]{Research Center for Magnetic and Spintronic Materials, National Institute for Materials Science, Tsukuba 305-0047, Japan}
\address[3]{PRESTO, Japan Science and Technology Agency, Saitama 332-0012, Japan}

\author[2,4]{Zixi Chen}
\address[4]{Graduate School of Pure and Applied Science, University of Tsukuba, Tsukuba 305-8571, Japan}
\author[2]{Jian Wang}
\author[2]{Weinan Zhou}
\author[2]{Kushwaha Varun}
\author[5]{Kenji Ueda}[orcid=0000-0001-7450-5501]
\address[5]{Department of Crystalline Materials Science, Graduate School of Engineering, Nagoya University, Nagoya 464-8603, Japan}
\author[6]{Shinya Yamada}
\address[6]{Center for Spintronics Research Network, Graduate School of Engineering Science, Osaka University, Toyonaka 560-8531, Japan}
\author[6]{Kohei Hamaya}
\author[2,4]{Kazuhiro Hono}[orcid=000-0001-7367-0193]
\cortext[cor1]{Corresponding authors}

\begin{abstract}
Inverse Heusler alloy Mn$_2$CoAl thin films, known as a spin-gapless semiconductor (SGS), grown by three different methods: ultra-high vacuum magnetron spattering, Ar-ion beam assisted sputtering, and molecular beam epitaxy, are investigated
by comparing their electric transport properties, microstructures and atomic-level structures.
Of the samples, the Mn$_2$CoAl thin film grown by MBE consists of Mn- and Co-rich phases,
the structures of which are determined to be the $L2_1B$-type and 
disordered $L2_1$-type, respectively, according to anomalous XRD analysis.
None of them forms the $XA$-type structure expected for SGS Heusler alloy,
although they all show SGS characteristics.
We suggest, to validate SGS characteristics,
it is necessary to extract not only magnetic and electric transport properties but also information about microstructures
and atomic-scale structures of the films including defects such as atomic swap.
\end{abstract}




\maketitle
\section{Introduction}
Heusler alloys have attracted much attention over a century
since the discovery of the first ferromagnetic copper-manganese-based alloys by Heusler in 1903~\cite{Felser2016},
due to its unique and diverse nature such as
ferromagnetism,
thermoelectronics~\cite{Hamzic1981,Nishino1997,Mikami2009},
magnetocaloric effect~\cite{Webster1984},
and shape memory effect~\cite{Sutou2004,Krenke2005}.
Especially, its half-metal property, i.e., conduction electrons that are 100\% spin-polarized
due to a gap at the Fermi level ($E_{\rm{F}}$) in the minority spin channel
with a finite density of states at the $E_{\rm{F}}$ for the majority spin channel~\cite{Groot1983},
is highly desirable for ferromagnetic (FM) materials in spintronic devices.
Among various potential half-metals, Co-based Heusler alloys are particularly promising as the FM electrodes of
magnetoresistive devices since most of them show half-metallicity~\cite{Kandpal2007}
and high Curie temperatures ($T_{\rm{c}}$)~\cite{Kuebler1983}.

Recently, a special class of semiconductor, so-called spin-gapless semiconductor (SGS), was proposed
using ab-initio calculations~\cite{Wang2008},
and SGS-like behaviors have been reported experimentally for the Mn-based Heusler compound Mn$_2$CoAl~\cite{Ouardi2013} that has high $T_{\rm{c}}$ like Co-based ones~\cite{Kuebler1983}.
SGS, in which the conduction and valence band edges meet at $E_{\rm{F}}$ and 
there is no gap for one spin channel while there is a finite gap in another spin channel~\cite{Wang2008},
gives rise to combined properties of a half-metal and semiconductor.
Therefore, SGSs have additional attractive properties other than 100\% spin-polarization such as
tunability of $E_{\rm{F}}$ by external electronic magnetic fields and carrier doping\cite{Wang2008},
which lead to easy switching between electrons and holes modes.
Thus, SGSs can provide highly spin polarized channels with tunable magnetic properties,
which is promising as highly efficient spin injectors.

The Mn$_2$CoAl compound has high Curie temperature (ca. 720 K),
small Seebeck coefficient ($< 1 \mu$ VK$^{-1}$),
low carrier concentration ($\sim 10^{17}$ cm$^{-3}$),
low conductivity with semiconducting temperature dependence ($\sim 10^3$ Scm$^{-1}$ ),
and positive magneto-resistance (MR) at low temperatures (< 100 K)~\cite{Ouardi2013},
which are considered as characteristic magnetic and electric transport properties of SGS.
However,
the recent work using
scanning transmission electron microscope (STEM) with energy dispersive X-ray spectroscopy(EDS)
and synchrotron X-ray diffraction (XRD) reported that
bulk Mn$_2$CoAl tends to show phase separation in the equilibrium state~\cite{Xu2019},
suggesting that a stoichiometric Mn$_2$CoAl compound is thermodynamically unstable.
For potential application of the material in devices,
Mn$_2$CoAl has been studied in the thin-film form as well~\cite{Jamer2013};
however, the SGS features of Mn$_2$CoAl thin films largely depend on fabrication conditions~\cite{Xu2014,Sun2016,Chen2018,Arima2018}.
This is considered to be due to the presence of various anti-site lattice defects in the thin films,
which causes different local symmetry that destroy the SGS character~\cite{Galanakis2014}.
Of the many papers on SGSs, there have been few reports on microstructures and atomic-scale structures,
despite this information is critical to validate their SGS characteristics.
In this respect, laboratory-based XRD is not sufficient to distinguish structural differences
in Heusler alloys due to the nearly identical atomic scattering factors of the constituent elements,
e.g., Co and Mn at the Cu K-absorption edge.
Therefore, more in-depth structural evaluations of SGS materials are essential
to understand their magnetic and transport properties.
Synchrotron anomalous XRD is a solution to address this issue,
which offers an element-specific analysis of atomic-level structures
in each crystalline phase and specific crystallographic sites within the phase,
i.e. the evaluation of the degree of order.

In this study, in addition to electric transport properties,
the microstructures and atomic-level structures in Mn$_2$CoAl thin films grown by magnetron sputtering (MS),
ion-beam assisted sputtering (IBAS),
and molecular beam epitaxy (MBE) are investigated.
%
Of the samples, the Mn$_2$CoAl thin film grown by MBE consists of Mn- and Co-rich phases,
the structures of which are determined to be the $L2_1B$-type and disordered $L2_1$-type
in the {\it Strukturbericht} designation~\cite{Ewald1931,Mehl2019}, respectively,
according to synchrotron anomalous XRD.
Although all the films show SGS characteristics,
we find none of them form the $XA$-type structure required for the SGS band structure of Mn$_2$CoAl.

\section{Experiments}
Mn$_2$CoAl thin films were prepared using three different growth methods as described below. The first sample was grown by conventional ultra high vacuum (UHV) MS system using a sintered Mn$_{50}$Co$_{25}$Al$_{25}$ alloy target. First, a MgO(001) single crystalline substrate was installed in a UHV chamber (base pressure is about $5\times10^{-7}$ Pa) and cleaned by Ar ion bombardment. Then, a 50-nm-thick Mn$_2$CoAl film was grown by direct-current (DC) magnetron sputtering at the elevated substrate temperature of 600$^\circ$C.
The second Mn$_2$CoAl film was grown by IBAS. A 90-nm-thick Mn$_2$CoAl film was grown from the Mn$_2$CoAl target
on a MgAl$_2$O$_4$ (001) substrate at the substrate temperature of 400$^\circ$C.
The Ar ion density from an Ar assist gun during deposition was $\sim$10$^{13}$ cm$^{-2}$ for the Ar beam and acceralator voltage of 100 V and 400 V, respectively.
This film was grown by the same IBAS system with almost the same deposition condition and procedure with the Mn$_2$CoAl film reported in a previous study~\cite{Ueda2017}.
The third Mn$_2$CoAl film was grown on a MgAl$_2$O$_4$ (001) substrate by MBE. First, the surface of MgAl$_2$O$_4$ substrate was cleaned by the heat treatment at 600$^\circ$C. After cooling down the substrate temperature to 300$^\circ$C, a 25-nm-thick Mn$_2$CoAl film was grown by co-evaporating Mn, Co and Al sources using Knudsen cells.

For screening the crystal quality in the three Mn$_2$CoAl films, these films were investigated by laboratory-based XRD with Cu K$\alpha$ radiation.
Longitudinal resistivity $\rho_{\rm{xx}}$ was measured by a DC four-probe method. Transverse resistivity $\rho_{\rm{xy}}$ was measured by flowing DC current and applying the magnetic field perpendicular to the film.
Thin samples for STEM-EDS were prepared by the focused ion beam machine, FEI, Helios G4 UX.
High-angle annular dark-field (HAADF) STEM images were acquired
with a probe forming aberration corrector operated at 200 kV (FEI, Titan G2 80-200).

Synchrotron anomalous XRD was carried out on the beamline BL13XU~\cite{Sakata2003} at SPring-8.
Anomalous XRD has following advantages: distinguishability of neighboring elements, sensitivity to specific crystalline phases and occupancy sites adequate for the evaluation of high entropy materials such as Heusler alloys.
Intense hard X-rays from the in-vacuum undulator and the six-circle diffractometer are available on the beamline.
After an asymmetric-cut Si111 double-crystal monochromator~\cite{Tajiri2019}, two mirrors are located for rejecting higher harmonics from the undulator.
The double slits configuration with a YAP scintillation detector in a $\theta$-2$\theta$ scan was employed
in the anomalous XRD measurements, which were performed with an out-of-plane geometry.
The X-ray energies near the Co (7.709 keV) and Mn (6.539 keV) $K$-absorption edges were used.
%
\section{Results}
\subsection{Laboratory XRD, micro structures and transport properties}
\begin{figure}
\centering
\includegraphics[width=0.9\linewidth]{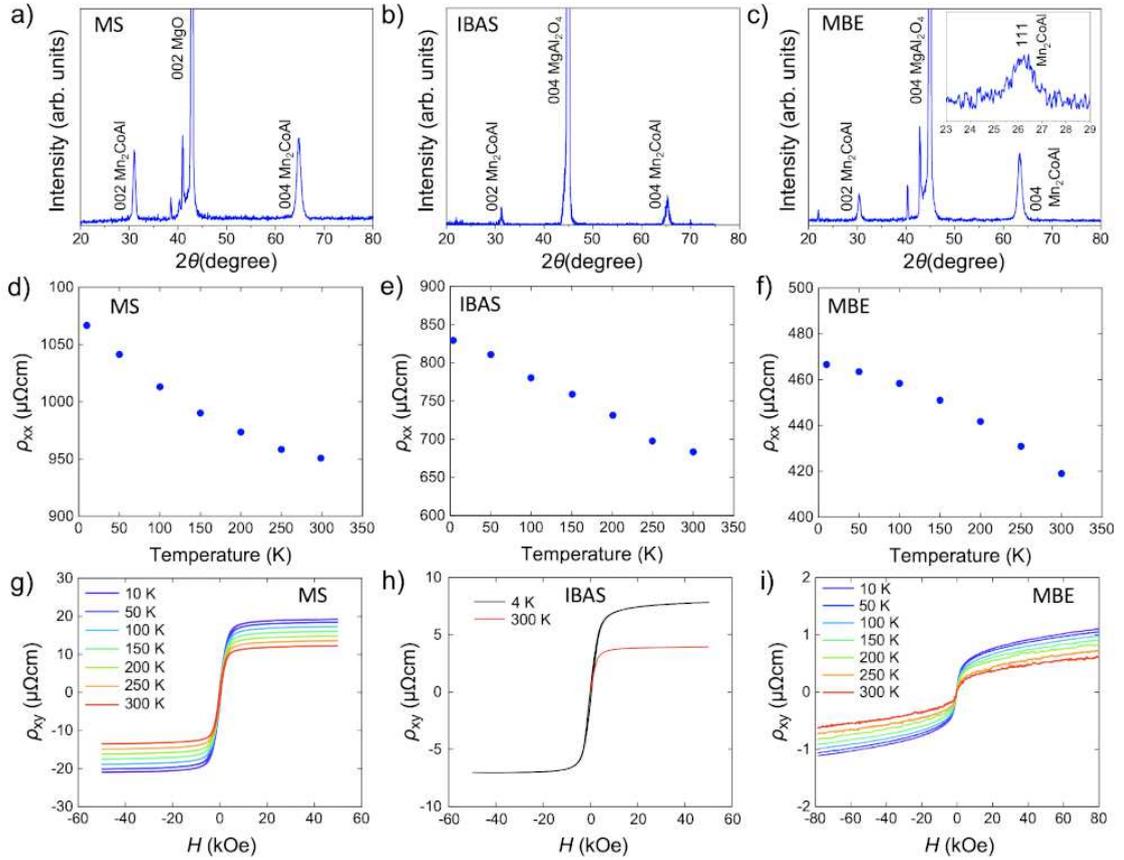}
\vspace{2.0em}
\caption{
(a)--(c) Laboratory-based XRD patterns,
(d)--(f) temperature dependence of longitudinal resistivities $\rho_{xx}$,
(g)--(i) magnetic-field dependence of transverse resistivities $\rho_{xy}$
on the Mn$_2$CoAl films grown by MS (left), IBAS (middle), and MBE (right), respectively.
}
\label{FIG:MCAProperty}
\end{figure}

Figs.~\ref{FIG:MCAProperty}(a)--(c) show the XRD patterns for the three Mn$_2$CoAl films grown by MS, IBAS and MBE, respectively. In these out-of-plane $\theta$-2$\theta$ scan profiles, we found only two peaks arising from the Mn$_2$CoAl film at around 31--32$^\circ$ and 64--66$^\circ$. The positions of these two peaks agree well with the peak positions of 002 superlattice and 004 fundamental peaks expected
in the $XA$- or $L2_1$-type structure.
We also measured a profile by tilting the film normal by 54.7$^\circ$ to see the 111 superlattice peak that mainly originates from the ordering between Mn on B-site and Al on D-site in the $XA$/$L2_1$ structure. A clear peak was observed at around 26$^\circ$ only in the Mn$_2$CoAl film grown by MBE as shown in the inset of Fig.~\ref{FIG:MCAProperty}(c).
Therefore, these XRD results suggest the formation of a $XA$- or $L2_1$-type structure in the MBE-Mn$_2$CoAl film and those with Mn(B) and Al(D) disorder in MS- and IBAS-Mn$_2$CoAl films.

The temperature dependences of the longitudinal resistivity $\rho_{\rm{xx}}$ for these three Mn$_2$CoAl films are shown in Figs.~\ref{FIG:MCAProperty}(d)--(f). $\rho_{\rm{xx}}$ in all films increases with decreasing temperature, which is usually interpreted as the semiconducting property of SGS.
Figs.~\ref{FIG:MCAProperty}(g)--(i) show the magnetic field dependence of transverse resistivity $\rho_{\rm{xy}}$ measured at different temperatures for the three Mn$_2$CoAl films. The anomalous Hall conductivity $\sigma_{\rm{AHE}}$ evaluated from these Hall curves at low temperature is 15, 10 and 4 Scm$^{-1}$ for the MS, IBAS and MBE-Mn$_2$CoAl films, respectively. Compared with the $\sigma_{\rm{AHE}}$ reported in the bulk Mn$_2$CoAl, 22 Scm$^{-1}$ at 2 K, by Ouardi et al~\cite{Ouardi2013}, these $\sigma_{\rm{AHE}}$ are comparably small, which also supports the SGS-like nature of these films.

\begin{figure}
\centering
\includegraphics[width=0.9\linewidth]{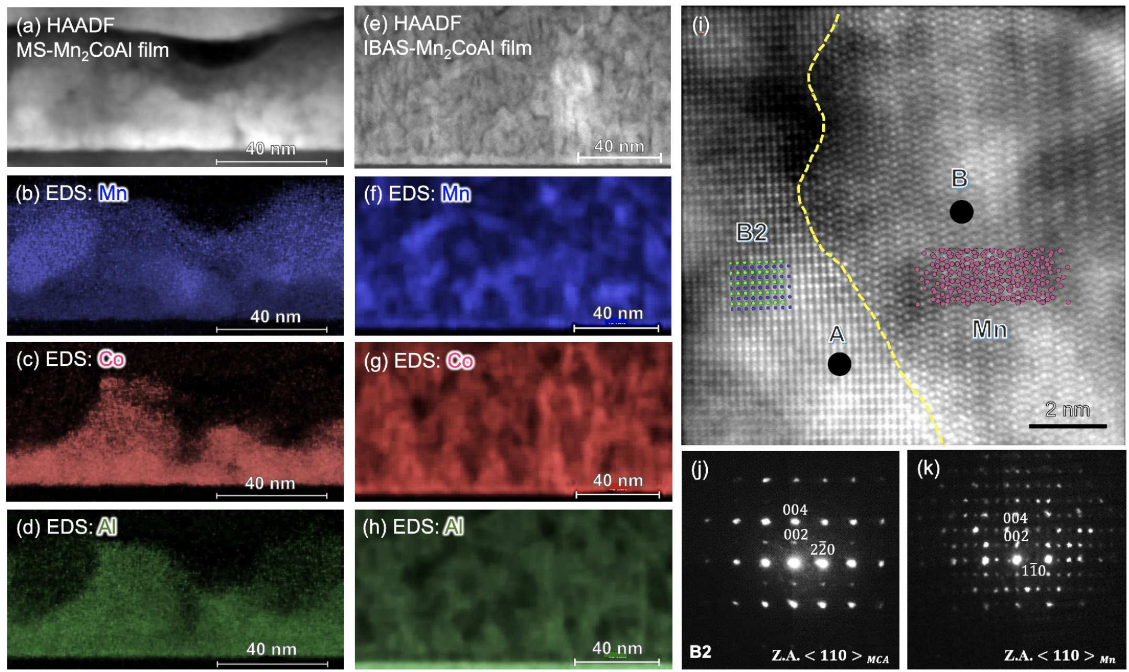}
\vspace{2.0em}
\caption{
HAADF-STEM and EDS mapping images of Mn, Co and Al for the MS-Mn$_2$CoAl film (a-d) and the IBAS-Mn$_2$CoAl film (e-h), respectively. (i) High resolution STEM image of the IBAS-Mn$_2$CoAl film, (j)(k) nano-beam diffraction images taken from the spot A and B shown in (i), respectively.
}
\label{FIG:HAADF}
\end{figure}

However, microstructure analyses for these films using STEM showed different aspects from what we expected from the above-mentioned XRD and transport properties. The HAADF-STEM and EDS mapping for the MS-Mn$_2$CoAl film show clear phase separations with Mn and Co as shown in Figs.~\ref{FIG:HAADF}(a)-(d).
This phase separation is naturally understood from the phase diagram of the MnCoAl in which there is no thermally equilibrium Mn$_2$CoAl phase at 600$^{\circ}$C but CoAl and Mn phases coexist~\cite{Kainuma1998}.
Since CoAl phase is expected to have a $B2$-ordered structure with the lattice constant of nearly half of that for Mn$_2$CoAl, the diffraction peaks appear at around 32 and 64$^{\circ}$ in the XRD pattern (Fig.~\ref{FIG:MCAProperty}(a)), that were regarded as 002 and 004 peaks of Mn$_2$CoAl.
However, these correspond to 001 and 002 peaks of $B2$-CoAl, respectively.
Although the origins for semiconducting temperature dependence of $\rho_{\rm{xx}}$ and small $\rho_{\rm{xy}}$ in the MS-Mn$_2$CoAl film are unclear,
it is inferred that the CoAl phase with a small amount of Mn dissolution might be ferromagnetic,
and this may be attribured to the anomalous Hall effect.

In the IBAS-Mn$_2$CoAl film, the phase separation is much weaker than the MS-Mn$_2$CoAl film as shown in Figs.~\ref{FIG:HAADF}(e)-(h),
which is explained by relatively lower temperature of the substrate during deposition
compared to that of the MS-Mn$_2$CoAl film thanks to an assists from ion-beam bombardment.
On the other hand, there is also Mn precipitates in the IBAS-Mn$_2$CoAl film that was clearly confirmed by high-resolution STEM and nano-beam diffraction images from the corresponding region of the Mn precipitates (Figs.~\ref{FIG:HAADF}(i)-(k)). The composition analysis based on EDS mapping indicates that there are regions with nearly Mn:Co:Al = 2:1:1, together with $B2$ ordered structure in the IBAS-Mn$_2$CoAl film, which might contribute to observed SGS-like transport properties~\cite{Ueda2017}.

\begin{figure}
\centering
\includegraphics[width=0.7\linewidth]{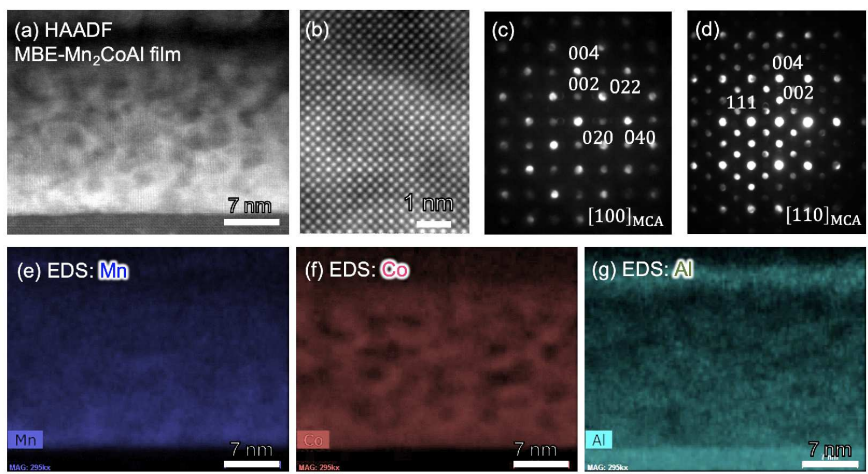}
\vspace{2.0em}
\caption{
(a) HAADF-STEM image, EDS mapping images of (e) Mn, (f) Co, and (g) Al for the MBE-Mn$_2$CoAl film. (b) High resolution STEM image of the MBE-Mn$_2$CoAl film, (c)(d) nano-beam diffraction images taken from the [100] and [110] directions of the MBE-Mn$_2$CoAl film, respectively.
}
\label{FIG:HAADF_MBE}
\end{figure}

In contrast to the MS- and IBAS-Mn$_2$CoAl films, the MBE-Mn$_2$CoAl film shows almost homogeneous microstructure as shown in Fig.~\ref{FIG:HAADF_MBE}(a).
High-resolution STEM and nano-beam diffraction images show only Heusler phase without a Mn precipitation (Figs.~\ref{FIG:HAADF_MBE}(b)-(d)). The 111 diffraction spots are clearly detected in the nano-beam diffraction taken from the [110] direction of the Mn$_2$CoAl film (Fig.~\ref{FIG:HAADF_MBE}(d)), suggesting the formation of $XA$ or $L2_1$-phases. It is expected that the high-quality Mn$_2$CoAl film was realized by a thermally non-equilibrium growth of a Mn$_2$CoAl film due to a low temperature MBE technique~\cite{Arima2018}.
%
In addition, careful analysis of EDS mapping of Mn, Co and Al revealed 
that Co-rich and Mn-rich regions in 5 to 10 nm scale (Figs.~\ref{FIG:HAADF_MBE}(e)-(g)).
The area selective composition analysis for the Co-rich and Mn-rich regions averaging of total 20 areas for each estimated
that the average composition ratio in the Co-rich region is Mn:Co:Al = 40.8$\pm$2.6 : 43.9$\pm$5.1 : 15.3$\pm$3.6 and Mn-rich region is Mn:Co:Al = 54.0$\pm$2.3 : 21.7$\pm$3.3 : 24.3$\pm$3.3 (See Fig.S1 in supplementary file).
%
\subsection{Synchrotron anomalous XRD}
\begin{figure}
\centering
\includegraphics[width=0.4\linewidth]{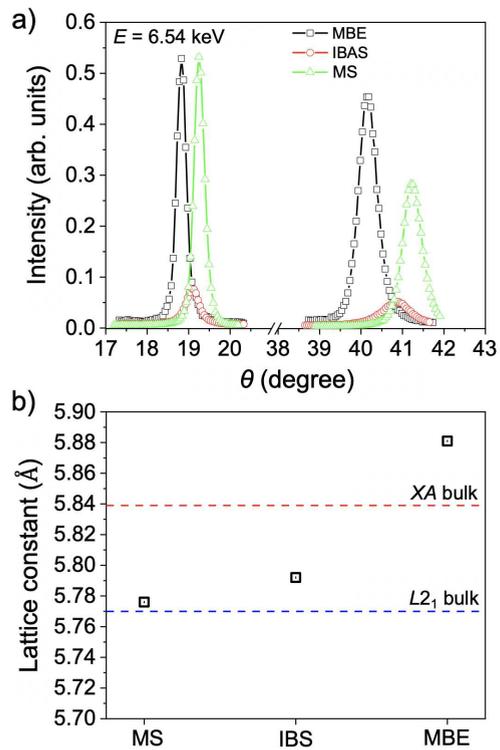}
\vspace{2.0em}
\caption{
(a) Synchrotron XRD rocking curves of the 002 and 004 reflections for the Mn$_2$CoAl thin films grown by MS, IBAS and MBE.
(b) Lattice constants of the Mn$_2$CoAl thin films determined by synchrotron XRD.
The lattice constants of bulk $XA$ and $L2_1$ structures, 0.584 nm and 0.577 nm, respectively, are shown as dotted lines.
}
\label{FIG:SynchrotronXRD}
\end{figure}

Fig.~\ref{FIG:SynchrotronXRD} shows the rocking curves of the 002 and 004 reflections
from the three Mn$_2$CoAl thin films grown by MS, IBAS, and MBE,
and the corresponding lattice constants derived from the curve peaks.
Among these samples, the film prepared by MBE most closely resembles to that of the bulk Mn$_2$CoAl
in terms of the $XA$ structure (0.584 nm).

\begin{figure}
\centering
\includegraphics[width=0.6\linewidth]{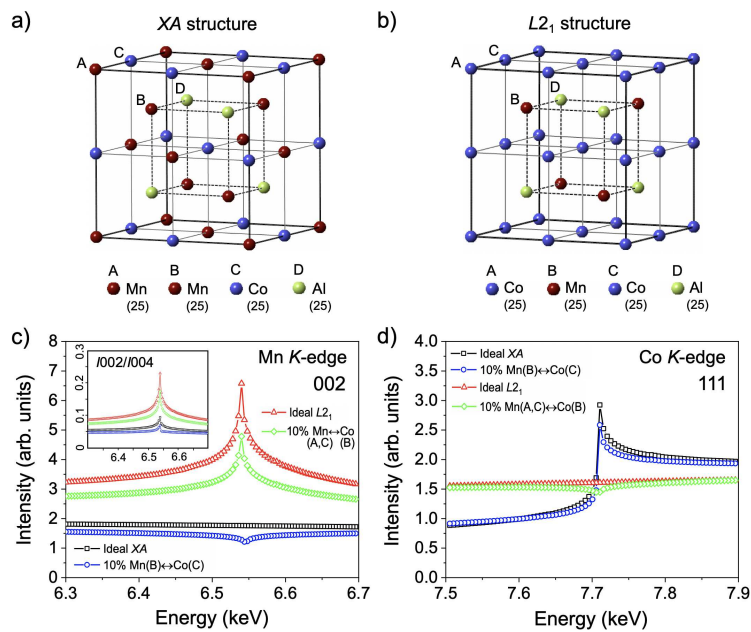}
\vspace{2.0em}
\caption{
Atomic structure models of (a) Mn$_2$CoAl in the $XA$ structure and (b) Co$_2$MnAl in the $L2_1$ structure.
Simulated anomalous XRD patterns of Mn$_2$CoAl $XA$ and Co$_2$MnAl $L2_1$ structures with disorder
in (c) $I_{002}$ near the Mn $K$-edge (inset of (c) shows $I_{002}$/$I_{004}$) and (d) $I_{111}$ near the Co $K$-edge.
}

\label{FIG:StructureAndIdealAXRD}
\end{figure}

The structure of the $XA$-type Mn$_2$CoAl inverse Heusler compound with space group $F\bar{4}3m$
is similar to that of the $L2_1$-type ($Fm\bar{3}m$) full Heusler compound Co$_2$MnAl~\cite{Galanakis2014},
as shown in Fig.~\ref{FIG:StructureAndIdealAXRD} (a) and (b),
whereas only the $XA$-type structure is predicted to show the SGS properties.
Figs.~\ref{FIG:StructureAndIdealAXRD} (c) and (d) show
the calculations of anomalous XRD for ideal $XA$-Mn$_2$CoAl and $L2_1$-Co$_2$MnAl,
and their disordered structures.
We observed a peak in the 002 reflection and a dip in the 111 reflection for $L2_1$. On other hand, the ideal $XA$ shows a dip in 002 and a peak in 111. Therefore, we can easily distinguish the $XA$ and $L2_1$ structures by anomalous XRD.

\begin{figure}
\centering
\includegraphics[width=0.4\linewidth]{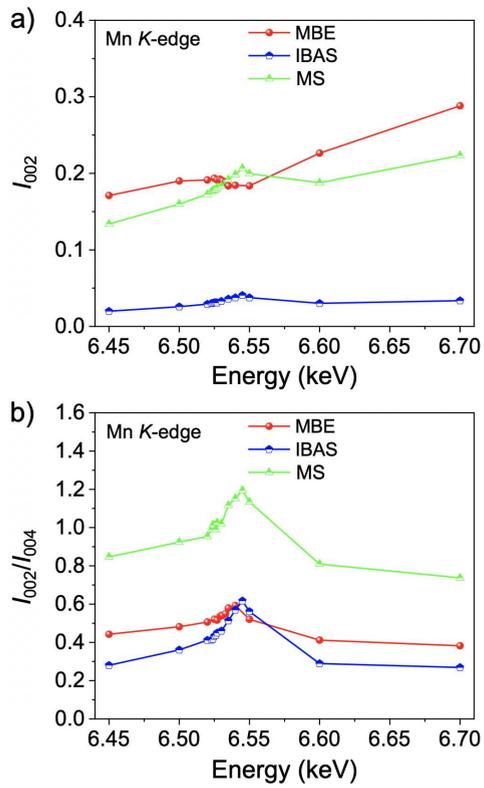}
\vspace{2.0em}
\caption{Experimental anomalous XRD patterns of Mn$_2$CoAl films grown by MBE, IBAS, and MS. (a) $I_{002}$  and (d) $I_{002}/I_{004}$ profiles near Mn $K$-edge.
}
\label{FIG:MS-IBAS-MCA-AXRD}
\end{figure}

Figs.~\ref{FIG:MS-IBAS-MCA-AXRD} (a) and (b)
show the experimentally measured anomalous XRD of the 002 superlattice reflection and
the intensity ratios $I_{002}$/$I_{004}$ of the 002 ($I_{002}$) and 004 fundamental reflection ($I_{004}$) near Mn $K$-edge, respectively, for the three Mn$_2$CoAl fabricated by MS, IBAS, and MBE.
The 002 reflection in the Mn$_2$CoAl film grown by MBE only shows a dip at the Mn K-edge
as expected from the calculated results for the $XA$ structure (Fig.~\ref{FIG:StructureAndIdealAXRD}(c)),
while the trends in $I_{002}$/$I_{004}$ for all Mn$_2$CoAl films are similar to those calculated for both the $XA$ and $L2_1$ structures,
as shown in the inset of Fig.~\ref{FIG:StructureAndIdealAXRD}(c).
According to these results, only the MBE-Mn$_2$CoAl film is expected to have a plausible $XA$ structure, but the others are not. Therefore, we further carried out a comprehensive atomic-level structural investigation on the MBE-Mn$_2$CoAl film. 
%

In the following structure analysis, the compositions for the two regions, estimated by STEM-EDS mapping mentioned above,
were set to adequate round numbers to simplify the analysis without losing accuracy as follows: Mn:Co:Al = 40:45:15 for the Co-rich region and Mn:Co:Al = 55:25:20 and 55:20:25 for the Mn-rich region
by considering ambiguity of the estimation.
We assumed that the above two regions (Co-rich and Mn-rich regions) are equally distibuted in the MBE-Mn$_2$CoAl film and consist of either the $XA$ or $L2_1$ structures as expected from the high-resolution STEM image in Fig.~\ref{FIG:HAADF_MBE}(b).
Under these assumptions,
we built all possible 13 candidates structural models for the Co-rich region (Table~\ref{tab:S1})
and 12 models for the Mn-rich region (Table~\ref{tab:S2})
by arranging site occupations by Mn, Co, and Al with 5\% intervals
within the $XA$ or $L2_1$ framework.
Each site A, B, C, and D in Tables~\ref{tab:S1} and \ref{tab:S2} corresponds to
the following Wyckoff positions, 4a (0, 0, 0), 4c (1/4, 1/4, 1/4), 4b (1/2, 1/2, 1/2), and 4d (3/4, 3/4, 3/4), respectively.

From the previous study on the similar Co-based Heusler system, Co$_2$(Mn$_{0.6}$Fe$_{0.4}$)Ge,
reporting the high-formation energy of Co-Ge than those of Co-Mn and Co-Fe confirmed by first principles calculations~\cite{Li2018},
it is known that Co atom is reluctant to occupy the Al site, unlikely showing X-Z disorder.
Therefore, we can reject model B-2 and B-6 with Co-rich $L2_1$ structure from the candidate in Table~\ref{tab:S1}
at the first glance.

\begin{figure}
	\centering
	\includegraphics[width=0.4\linewidth]{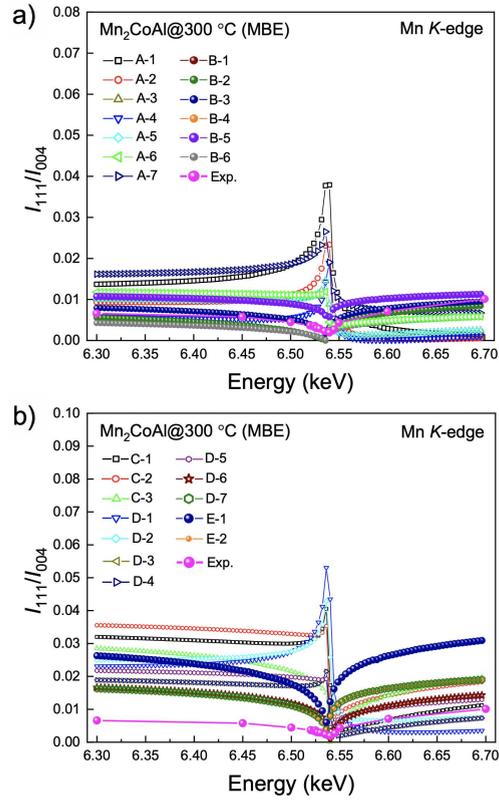}
	\vspace{2.0em}
	\caption{Calculated anomalous XRD profiles for the (a) Co-rich and (b) Mn-rich regions of a MBE-Mn$_2$CoAl thin film compared with experimentally obtained $I_{111}/I_{004}$ around the Mn $K$-edge.
	The calculations were performed based on $XA$ and $L2_1$ structure models in both the Co-rich and Mn-rich regions as shown in Tables.~\ref{tab:S1} and \ref{tab:S2}. }
	\label{FIG:AXRDCalculation}
\end{figure}

For the next step,
we calculated the X-ray energy dependence of diffraction intensities, $I_{111}$, $I_{002}$, and $I_{004}$, for all candidate models (25 models) in Tables \ref{tab:S1} and \ref{tab:S2} at around both the Co $K$-absorption edge and Mn $K$-absorption edge by utilizing the analysis software for XRD (SISReX) developed by one of the authors~\cite{Tajiri2020}.
Thus, totally 150 profiles were investigated. 
For example, Fig.~\ref{FIG:AXRDCalculation} shows the experimental and calculated energy dependence of the integrated intensity ratio, $I_{111}$/$I_{004}$ for both the Co-rich and Mn-rich regions. 
By comparing calculations with the experimental results of $I_{002}$, $I_{111}$, $I_{002}/I_{004}$, and $I_{111}/I_{004}$,
the best models for the Co-rich and Mn-rich regions were selected by using the figure of merit (FOM) that is derived from the summation of all degree of agreements according to the following first criteria:
+1: similar dip or peak shape;
-1: opposite profile;
0: no peak or dip shape.
After the first screening using the FOM,
13 models were reduced to 2 models (A-6 and B-5) for the Co rich region and
12 models were reduced to 4 models (D-6, D-7, E-1 and E-2) for the Mn-rich region
(See the Supplemental file for in detail).

Then, as shown in Table~\ref{tab:S5}, we consider 8 possible combinations to construct mixed structures
from the selected models of the Co-rich (2 models) and Mn-rich (4 models) regions.  
As shown in Table~\ref{tab:S5}, we introduced the additive criteria for deeper evaluations of the mixed 
structures:
+1; a calculated profile has similar shape and peak/dip height with that experimentally obtained,
 0; similar shape but different peak/dip height,
-1; opposite profile or asymmetrical,
resulting in that the two combinations (B-5\&D-7 and B-5\&E-2) are plausible.
(See the Supplemental file for in detail).
Now we notice that models Mn-rich $XA$ D-7 and $L2_1$ E-2 are indeed identical in its crystallographic definitions.
Finally, we obtained the best mixed structure consists of models B-5 and D-7,
which are attributed to disordered $L2_1$ structure in the Co-rich region and
$L2_1B$ structure in the Mn-rich region, respectively. 
Throughout the above analysis,
structural models which have $XA$-like anomalous XRD profiles as shown in Figs.~\ref{FIG:StructureAndIdealAXRD}(c) and (d)
tend to get low score than those with $L2_1$-like profiles. These tendencies also support the structures of the MCA-Mn$_2$CoAl film are not the $XA$-type.

\begin{figure}
\centering
\includegraphics[width=0.6\linewidth]{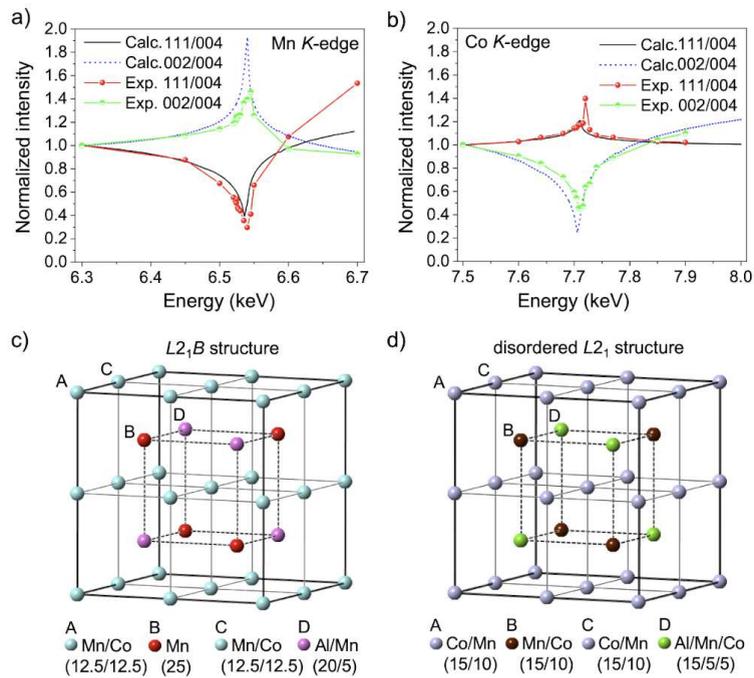}
\vspace{2.0em}
\caption{
Anomalous XRD profiles of $I_{111}/I_{004}$  and $I_{002}/I_{004}$ near the (a) Mn $K$-edge and (b) Co $K$-edge of the Mn$_2$CoAl film. Points with lines show the experimental results, while solid and dotted lines show calculated profiles based on the two phase models with the obtained atomic compositions and occupancies.
$L2_1B$-type and the disordered $L2_1$-type structures were determined for the Mn-rich and Co-rich phases, respectively,
by anomalous XRD of the Mn$_2$CoAl film grown by MBE.
}
\label{FIG:FinalAXRDAndStructure}
\end{figure}

Fig.~\ref{FIG:FinalAXRDAndStructure} summarizes the final results of anomalous XRD, showing experimental and calculated anomalous XRD profiles and the final structure models of the MBE-Mn$_2$CoAl thin film.
We determined that $L2_1B$ and disordered $L2_1$ crystal structures could be assigned to the  Mn-rich and Co-rich regions of the MBE-Mn$_2$CoAl film, respectively.
The calculated anomalous XRD profiles at the Mn and Co K-edges were in good agreement with the experimental results. 
As a result, none of the three films including the MBE-Mn$_2$CoAl thin film was the $XA$-type structure,
contrary to our expectations from their SGS-like transport properties.

\section{Discussion}
So far, the influence of various kinds of disorder in Heusler alloy on its functionalities have been discussed~\cite{Felser2016}.
For example, although it is no doubt that half-metal is the ideal candidate materials for spintronic applications, the spin polarizations of Heusler alloys
are degraded by structural disorder, surface/interface stoichiometry, thermal fluctuations, etc.
Regarding Co-based full Heusler alloys Co$_2$YZ, forming the $L2_1$ structure,
its spin polarization is rapidly degraded by introducing the $D0_3$-type structural disorder,
while the $B2$-type disorder (Y-Z disorder) hardly affect its spin polarization~\cite{Miura2004,Miura2006}.
Recent experiments have demonstrated that the bulk spin polarization of the half-metal Co$_2$Mn(Ge$_{0.75}$Ga$_{0.25}$) crystalline film is enhanced through suppressing the formation of Co$_{\mathrm{Mn}}$ antisites (Mn site occupied by Co atom) by increasing Mn composition rate~\cite{Li2016},
and even in the polycrystalline Co$_2$(Mn$_{0.6}$Fe$_{0.4}$)Ge film,
its current-perpendicular-to-plane giant magnetoresitive outputs are enhanced by improving the $B2$-order,
i.e., eliminating Co$_{\mathrm{Mn}}$ and Co$_{\mathrm{Fe}}$ antisites~\cite{Li2018}.
%
In contrast to the robustness of half-metallicity against several types of disorder and defects mentioned above,
the electronic structure of SGS is predicted to be very sensitive to any disorder and defect.
While plenty of Heusler alloys have been predicted to exhibit the SGS behavior~\cite{Skaftouros2013,Jakobsson2015,Galanakis2016}
even in quaternary compounds~\cite{Bainsla2015,Bainsla2015a,Yamada2019},
defects such as atomic swaps between sites with different local symmetry easily
destroy a pseudo-gap of one spin-channel in the SGS, thus, only a half-metallic gap in another spin channel remains
~\cite{Galanakis2014,Kudrnovsky2013,Choudhary2016}.
Although the Mn$_2$CoAl thin films prepared using three different growth methods showed SGS characteristics,
we revealed that they do not have the $XA$-type structure required for the SGS band structure.
Without seeing microstructures and atomic-scale structures,
previous studies for Heusler-based SGS thin films claimed the property of SGS based on the semiconducting temperature dependence of resistivity
and small anomalous Hall conductivity, as observed in the Mn$_2$CoAl films.
However, these properties cannot be evidence for a formation of the SGS band structure at all.
We emphasize that evaluation of the aforementioned structural disorder experimentally
is critical to validate their SGS characteristics.
Synchrotron anomalous XRD definitely has addressed this issue
by revealing disordered structures in the MBE-Mn$_2$CoAl film at the atomic level,
i.e., disorders between close elements, Co and Mn, even in the phase separated Heusler alloy.

\section{Conclusion}
We have investigated the Mn$_2$CoAl thin films grown by MS, IBAS, and MBE
by comparing their electric transport properties, microstructures and atomic-level structures.
Of the samples, the Mn$_2$CoAl thin film grown by MBE consisted of Mn- and Co-rich phases,
the structures of which have been determined to be $L2_1B$-type and 
the disordered $L2_1$-type, respectively, according to anomalous XRD analysis.
None of them have formed the $XA$-type structure required for the SGS band structure of Mn$_2$CoAl,
although they all showed SGS characteristics.
The evaluation of various anti-site lattice defects in the films that destroy the SGS band structure
is critical to validate SGS characteristics.
Synchrotron anomalous XRD is one of the key solutions to address this issue.
%
\section{Acknowledgments}
The authors thank to Drs. T. Sasaki and X. Xu for their valuable discussion.
The synchrotron radiation experiments were performed on beamline BL13XU at SPring-8 with the approval of the JASRI (proposal Nos. 2017B1313, 2018A1231, 2018B0927, 2018B1162, 2018B1538, and 2018B2092). This work was partially supported by JSPS KAKENHI (grant Nos. 17H06152 and 18KK0111) and
PRESTO from the Japan Science and Technology Agency (No. JPMJPR17R5).
%
%
%

\bibliographystyle{elsarticle-num-names}

\bibliography{ActaMater_MCA_AXRD}

%
%
\clearpage
\clearpage
\begin{table}
	\centering
	\caption{\bf List of site occupations for structural models in the Co-rich region of the MBE-Mn$_2$CoAl thin film
with the composition of Mn:Co:Al = 40:45:15.}
	\begin{tabular}{cccccccccccccc}
		\hline
		\multicolumn{2}{c} {} & \multicolumn{2}{c} {A site} & & \multicolumn{2}{c} {B site} & &\multicolumn{2}{c} {C site} &&  \multicolumn{3}{c} {D site} \\
		\cline{3-4} \cline{6-7} \cline{9-10} \cline{12-14}
		Model No. & Structure & Mn & Co && Mn & Co && Mn & Co && Mn & Co & Al \\
		\hline
		   & Ideal $XA$   & 25 & 0 && 25 & 0 && 0 & 25 && 0 & 0 & 25 \\
		\hline
		A-1 & Co-rich $XA$ & 25 & 0 && 15 & 10 && 0 & 25 && 0 & 10 & 15 \\
		A-2 & Co-rich $XA$ & 20 & 5 && 20 & 5 && 0 & 25 && 0 & 10 & 15 \\
		A-3 & Co-rich $XA$ & 20 & 5 && 15 & 10 && 5 & 20 && 0 & 10 & 15 \\
		A-4 & Co-rich $XA$ & 15 & 10 && 25 & 0 && 0 & 25 && 0 & 10 & 15 \\
		A-5 & Co-rich $XA$ & 15 & 10 && 20 & 5 && 0 & 25 && 5 & 5 & 15 \\                        
		A-6 & Co-rich $XA$ & 15 & 10 && 15 & 10 && 5 & 20 && 5 & 5 & 15 \\                        
		A-7 & Co-rich $XA$ & 15 & 10 && 15 & 10 && 0 & 25 && 10 & 0 & 15 \\        
		\hline
		   & Ideal $L2_1$ & 0 & 25 && 25 & 0 && 0 & 25 && 0 & 0 & 25 \\
		\hline
		B-1 & Co-rich $L2_1$ & 5 & 20 && 25 & 0 && 5 & 20 && 5 & 5 & 15 \\
		B-2 & Co-rich $L2_1$ & 10 & 15 && 20 & 5 && 10 & 15 && 0 & 10 & 15 \\
		B-3 & Co-rich $L2_1$ & 2.5 & 22.5 && 25 & 0 && 2.5 & 22.5 && 10 & 0 & 15 \\
		B-4 & Co-rich $L2_1$ & 5 & 20 && 20 & 5 && 5 & 20 && 10 & 0 & 15 \\
		B-5 & Co-rich $L2_1$ & 10 & 15 && 15 & 10 && 10 & 15 && 5 & 5 & 15 \\
		B-6 & Co-rich $L2_1$ & 10 & 15 && 20 & 5 && 10 & 15 && 0 & 12.5 & 12.5 \\
		\hline
	\end{tabular}
	\label{tab:S1}
\end{table}

\clearpage
\begin{table}
	\centering
	\caption{\bf List of site occupations for structural models in the Mn-rich region of the MBE-Mn$_2$CoAl thin film
with compositions of Mn:Co:Al = 55:25:20 and 55:20:25.}
	\begin{tabular}{cccccccccccccc}
		\hline
		\multicolumn{2}{c} {} & \multicolumn{2}{c} {A site} & & \multicolumn{2}{c} {B site} & &\multicolumn{2}{c} {C site} &&  \multicolumn{3}{c} {D site} \\
		\cline{3-4} \cline{6-7} \cline{9-10} \cline{12-14}
		Model No. & Structure & Mn & Co && Mn & Co && Mn & Co && Mn & Co & Al \\
		\hline
		   & Ideal $XA$ & 25 & 0 && 25 & 0 && 0 & 25 && 0 & 0 & 25 \\
		\hline
		C-1 & Mn-rich $XA$ & 25 & 0 && 25 & 0 && 5 & 20 && 0 & 0 & 25 \\
		C-2 & Mn-rich $XA$ & 25 & 0 && 20 & 5 && 10 & 15 && 0 & 0 & 25 \\
		C-3 & Mn-rich $XA$ & 20 & 5 && 25 & 0 && 10 & 15 && 0 & 0 & 25 \\
		\hline
		D-1 & Mn-rich $XA$ & 25 & 0 && 25 & 0 && 0 & 25 && 5 & 0 & 20 \\
		D-2 & Mn-rich $XA$ & 25 & 0 && 20 & 5 && 5 & 20 && 5 & 0 & 20 \\
		D-3 & Mn-rich $XA$ & 25 & 0 && 20 & 5 && 10 & 15 && 0 & 5 & 20 \\
		D-4 & Mn-rich $XA$ & 20 & 5 && 25 & 0 && 5 & 20 && 5 & 0 & 20 \\
		D-5 & Mn-rich $XA$ & 20 & 5 && 20 & 5 && 10 & 15 && 5 & 0 & 20 \\
		D-6 & Mn-rich $XA$ & 15 & 10 && 25 & 0 && 10 & 15 && 5 & 0 & 20 \\
		D-7 & Mn-rich $XA$ & 12.5 & 12.5 && 25 & 0 && 12.5 & 12.5 && 5 & 0 & 20 \\
		\hline
		   & Ideal $L2_1$ & 0 & 25 && 25 & 0 && 0 & 25 && 0 & 0 & 25 \\
		\hline
		E-1 & Mn-rich $L2_1$ & 12.5 & 12.5 && 25 & 0 && 12.5 & 12.5 && 0 & 0 & 25 \\
		E-2 & Mn-rich $L2_1$ & 12.5 & 12.5 && 25 & 0 && 12.5 & 12.5 && 5 & 0 & 20 \\
		\hline
	\end{tabular}
	\label{tab:S2}
\end{table}

\clearpage
\begin{table}
	\centering
	\caption{\bf Figures of merit (FOM) for the mixed structure models, that are selected from both the Co-rich and Mn-rich regions, on the anomalous XRD results. The FOMs were derived from summation of all degree of agreements according to the following criteria:
+1; a calculated profile has similar shape and peak(dip) height(depth) with that experimentally obtained,
 0; similar shape but different peak(dip) height(depth),
-1; opposite profile or asymmetrical. }
	\begin{tabular}{ccccccccc}
		\hline
		Exp. peak  &  \multicolumn{8}{c} {Co-rich + Mn-rich } \\
		\cline{2-9}
		& A-6\&D-6 & B-5\&D-6 & A-6\&D-7 & B-5\&D-7 & A-6\&E-1 & B-5\&E-1 & A-6\&E-2 & B-5\&E-2 \\
		\hline
		I$_{111}$@Mn-$K$ & -1 & -1 &  1 &  1 &  0 &  0 &  1 &  1 \\
		
		I$_{111}$@Co-$K$ & -1 & -1 & -1 &  1 & -1 &  0 & -1 &  1 \\
		
		I$_{002}$@Mn-$K$ & -1 & -1 & -1 & -1 & -1 & -1 & -1 & -1 \\
		
		I$_{002}$@Co-$K$ &  1 &  1 &  1 &  1 &  0 &  0 &  1 &  1 \\
		
		I$_{111}$/I$_{004}$@Mn-$K$ & -1 & -1 &  0 &  1 &  1 & 0 & 0 & 1 \\
		
		I$_{111}$/I$_{004}$@Co-$K$ &  0 & -1 &  0 &  1 &  0 & 0 & 0 & 1 \\
		
		I$_{002}$/I$_{004}$@Mn-$K$ &  1 &  1 &  1 &  1 &  0 & 0 & 1 & 1  \\
		
		I$_{002}$/I$_{004}$@Co-$K$ &  0 &  0 &  0 &  0 &  1 & 1 & 0 & 0  \\
		\hline
		\bf {FOM } & -2 & -3 & 1 & \bf {5} & 0 & 0 & 1 & \bf {5}   \\
		\hline
	\end{tabular}
	\label{tab:S5}
\end{table}

\clearpage

\clearpage

\clearpage

\clearpage

\clearpage

\clearpage

\clearpage

\clearpage


\end{document}


\let\WriteBookmarks\relax
\def\floatpagepagefraction{1}
\def\textpagefraction{.001}
\shorttitle{}
\shortauthors{H. Tajiri {\it et~al.}}

\title [mode = title]{Supplementary File:\\ Structural insight using anomalous XRD into Mn$_2$CoAl Heusler alloy films grown by magnetron sputtering, IBAS and MBE techniques}


\author[1]{Hiroo Tajiri}
\cormark[1]
\ead{tajiri@spring8.or.jp}
\address[1]{Center for Synchrotron Radiation Research, Japan Synchrotron Radiation Research Institute, Hyogo 679-5198, Japan}

\author[1]{Loku Singgappulige Rosantha Kumara}[orcid=0000-0001-9160-6590]

\author[2,3]{Yuya Sakuraba}[orcid=0000-0003-4618-9550]
\address[2]{Research Center for Magnetic and Spintronic Materials, National Institute for Materials Science, Tsukuba 305-0047, Japan}
\address[3]{PRESTO, Japan Science and Technology Agency, Saitama 332-0012, Japan}

\author[2,4]{Zixi Chen}
\address[4]{Graduate School of Pure and Applied Science, University of Tsukuba, Tsukuba 305-8571, Japan}
\author[2]{Jian Wang}
\author[2]{Weinan Zhou}
\author[2]{Kushwaha Varun}
\author[5]{Kenji Ueda}[orcid=0000-0001-7450-5501]
\address[5]{Department of Crystalline Materials Science, Graduate School of Engineering, Nagoya University, Nagoya 464-8603, Japan}
\author[6]{Shinya Yamada}
\address[6]{Center for Spintronics Research Network, Graduate School of Engineering Science, Osaka University, Toyonaka 560-8531, Japan}
\author[6]{Kohei Hamaya}
\author[2,4]{Kazuhiro Hono}[orcid=000-0001-7367-0193]
\cortext[cor1]{Corresponding author}

\maketitle

\section{Composition analysis}
\begin{figure}[htbp]
	\centering
	\includegraphics[width=0.7\linewidth]{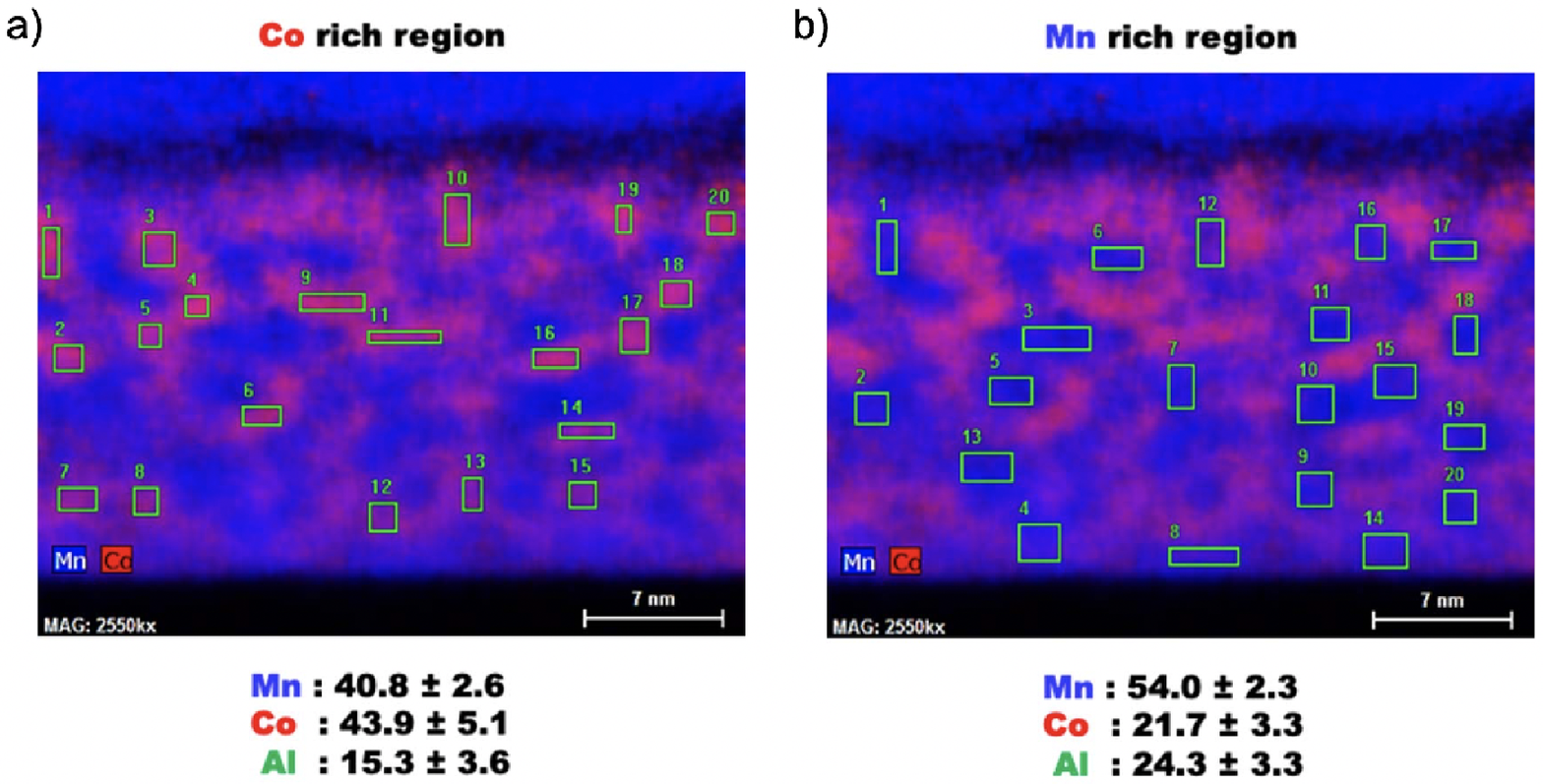}
	\vspace{2.0em}
	\caption{Area selective composition analysis of the MBE Mn$_2$CoAl film in (a) the Co-rich region and (b) the Mn-rich region using STEM-EDS mapping.}
	\label{FIG:S1}
\end{figure}

Figures~\ref{FIG:S1}(a) and (b) show the area selective composition analysis of the Mn$_2$CoAl thin film grown by molecular beam epitaxy (MBE).
Determined compositions of the MBE-Mn$_2$CoAl film in the Co-rich and Mn-rich regions are
Mn:Co:Al = 40.8$\pm$2.6 : 43.9$\pm$5.1 : 15.3$\pm$3.6, and 54.0  $\pm$2.3 : 21.7$\pm$3.3 : 24.3$\pm$3.3, respectively,
averaging compositions of total 20 areas for each.

\section{Analysis of anomalous XRD}
\subsection{Screening models in each region}
\begin{figure}[htbp]
	\centering
	\includegraphics[width=0.7\linewidth]{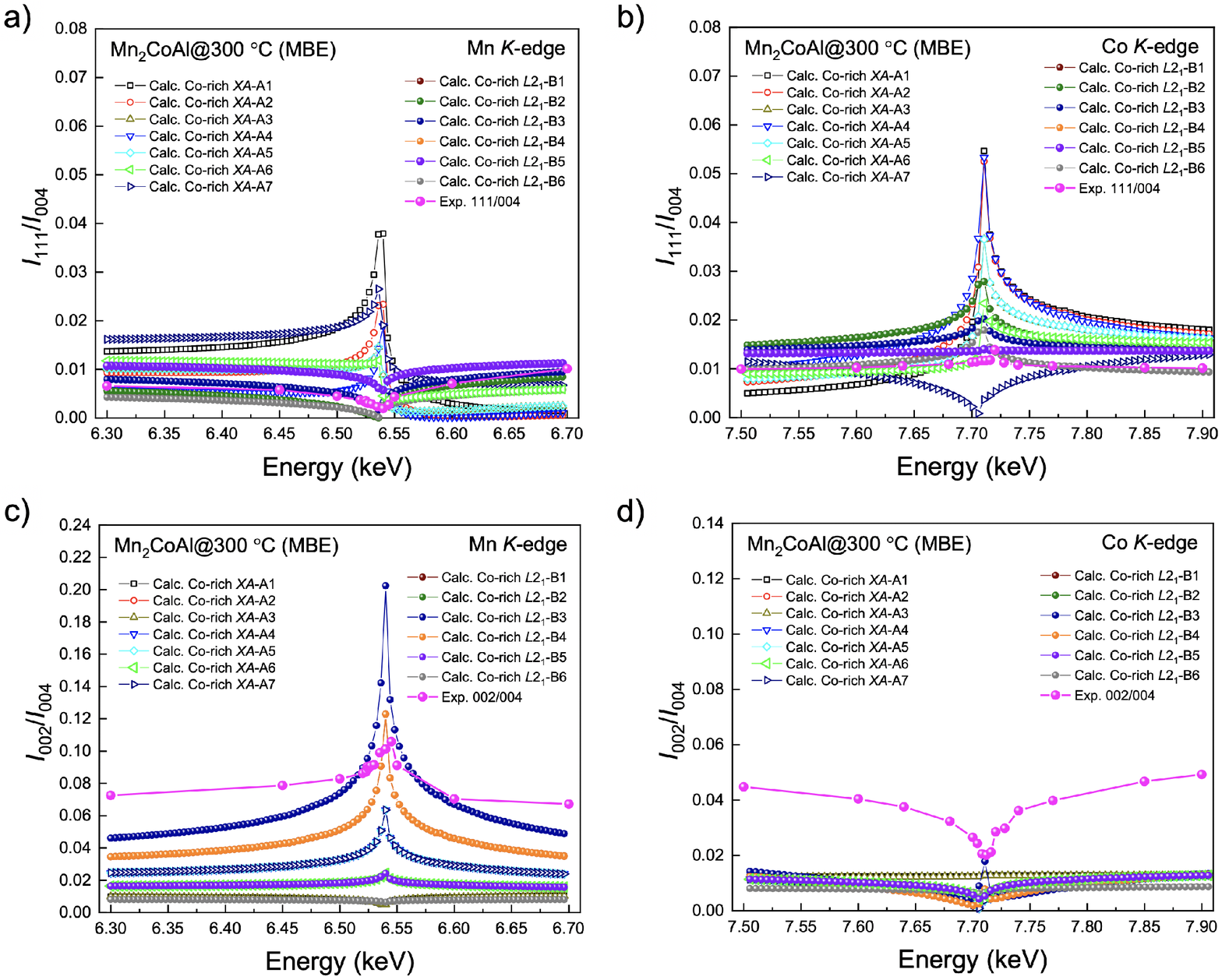}
	\vspace{2.0em}
	\caption{Calculated anomalous XRD results for the Co-rich region of Mn$_2$CoAl thin film compared with experimentally obtained (a) $I_{111}/I_{004}$ at Mn $K$-edge,  (b) $I_{111}/I_{004}$ at Co $K$-edge, (c) $I_{002}/I_{004}$ at Mn $K$-edge,  and (d) $ I_{002}/I_{004}$ at Co $K$-edge.  The calculations were performed based on $XA$ and $L2_1$ structure models in the Co-rich region.  }
	\label{FIG:S2}
\end{figure}

\begin{figure}[htbp]
	\centering
	\includegraphics[width=0.7\linewidth]{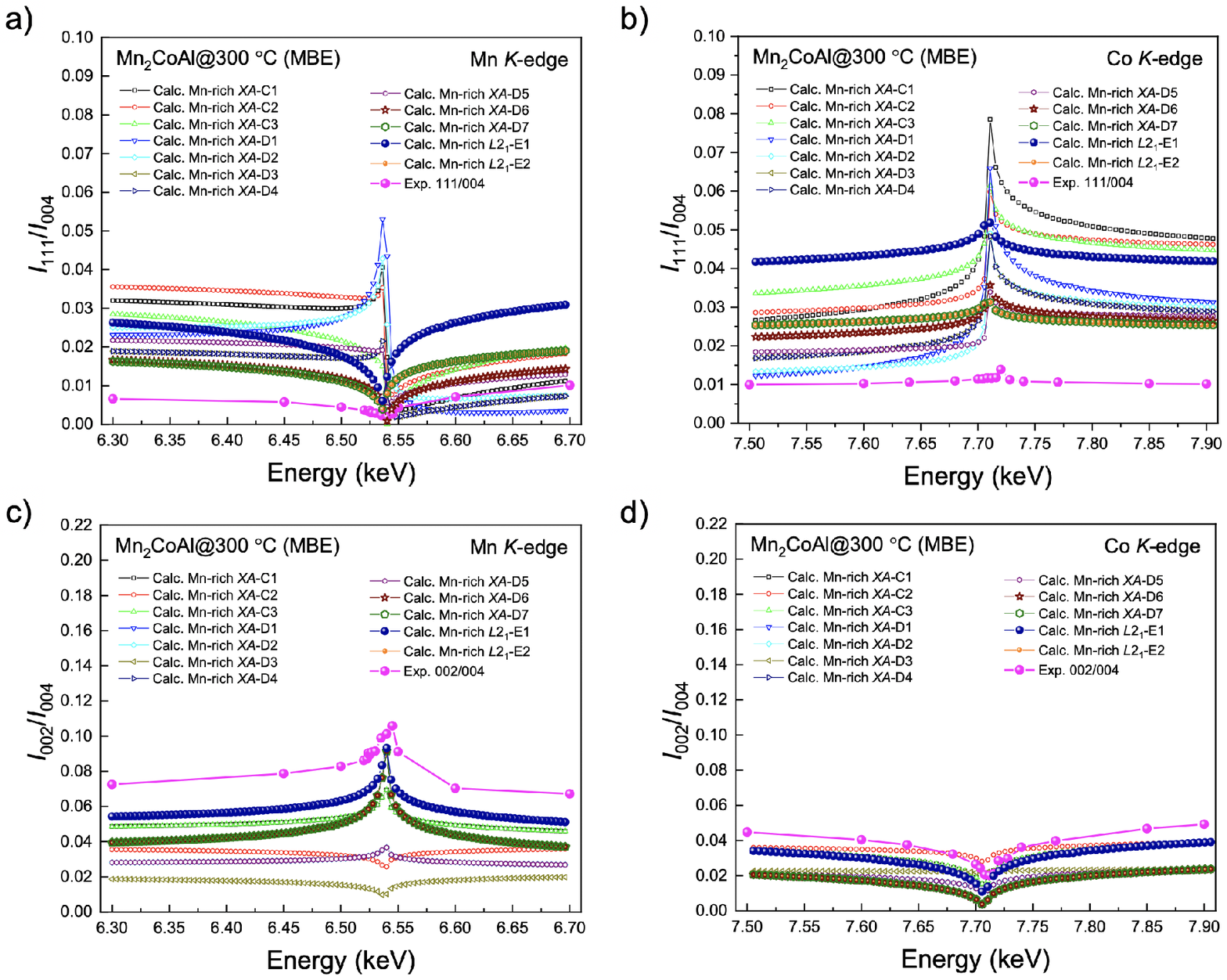}
	\vspace{2.0em}
	\caption{Calculated anomalous XRD results for the Mn-rich region of Mn$_2$CoAl thin film compared with experimentally obtained (a) $I_{111}/I_{004}$ at Mn $K$-edge,  (b) $I_{111}/I_{004}$ at Co $K$-edge, (c) $I_{002}/I_{004}$ at Mn $K$-edge,  and (d)$ I_{002}/I_{004}$ at Co $K$-edge.  The calculations were performed based on $XA$ and $L2_1$ structure models in the Mn-rich region. }
	\label{FIG:S3}
\end{figure}

Figures~\ref{FIG:S2} and \ref{FIG:S3} show the X-ray energy dependence of diffraction intensity ratios $I_{111}/I_{004}$ and $I_{002}/I_{004}$
for the structural models in Tables~1 (Co-rich region) and 2 (Mn-rich region) in the main text, respectively,
at around both the Co $K$-absorption edge (7.709 keV) and Mn $K$-absorption edge (6.539 keV),
in addition to the experimental data.
The integrated intensities at each X-ray energy were measured by rocking scans and corrected by the Lorentz factor, polarization factor (unity in our experiment), footprint factor, and absorption factor.
Synchrotron anomalous X-ray diffraction (XRD) was carried out on the beamline BL13XU~\cite{Sakata2003,Tajiri2019} at SPring-8.
Figures~\ref{FIG:S2}(a) and \ref{FIG:S3}(a) also appear in Figs.~7(a) and (b) in the main text, respectively.

Tables~\ref{tab:S3} and \ref{tab:S4} show
the figure of merit (FOM) that is derived from the summation of all degree of agreements according to the following first criteria:
+1: similar dip or peak shape;
-1: opposite profile;
0: no peak or dip shape,
by comparing calculations for the Co-rich region and Mn-rich region, respectively,
with the experimental results of $I_{002}$, $I_{111}$, $I_{002}/I_{004}$, and $I_{111}/I_{004}$.
Here, B-2 and B-6 models in Table~\ref{tab:S3} (Site occupations can be found in Table.~1 in the main text) were rejected
since excess amount of Co atom is not plausible to sit on D site in $L2_1$ structure~\cite{Li2018}.
After the first screening using the above FOM,
13 models were reduced to 2 models (A-6 and B-5) for the Co rich region and
12 models were reduced to 4 models (D-6, D-7, E-1 and E-2) for the Mn-rich region.

\subsection{Calculations for mixed structure}
\begin{figure}[htbp]
	\centering
	\includegraphics[width=0.7\linewidth]{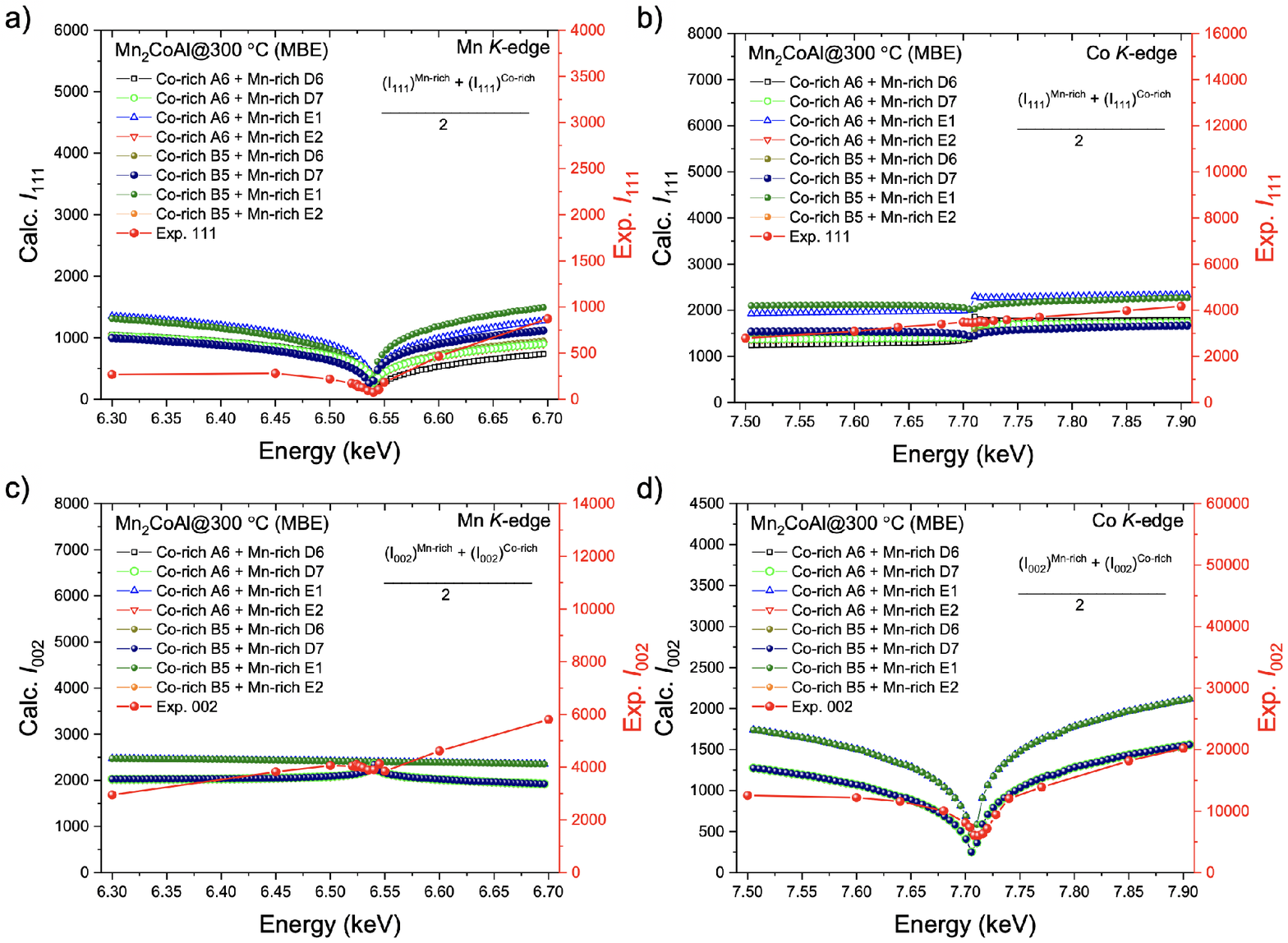}
	\vspace{2.0em}
	\caption{Calculated anomalous XRD for mixed structure compared with experimentally obtained (a) $I_{111}$ at Mn $K$-edge,  (b) $I_{111}$ at Co $K$-edge, (c) $I_{002}$ at Mn $K$-edge,  and (d) $ I_{002}$ at Co $K$-edge. The combinations were created using selected A-6 and B-5 models in the Co-rich region and another four D-6, D-7, E-1 and E-2 models in the Mn-rich region.}
	\label{FIG:S4}
\end{figure}

\begin{figure}[htbp]
	\centering
	\includegraphics[width=0.7\linewidth]{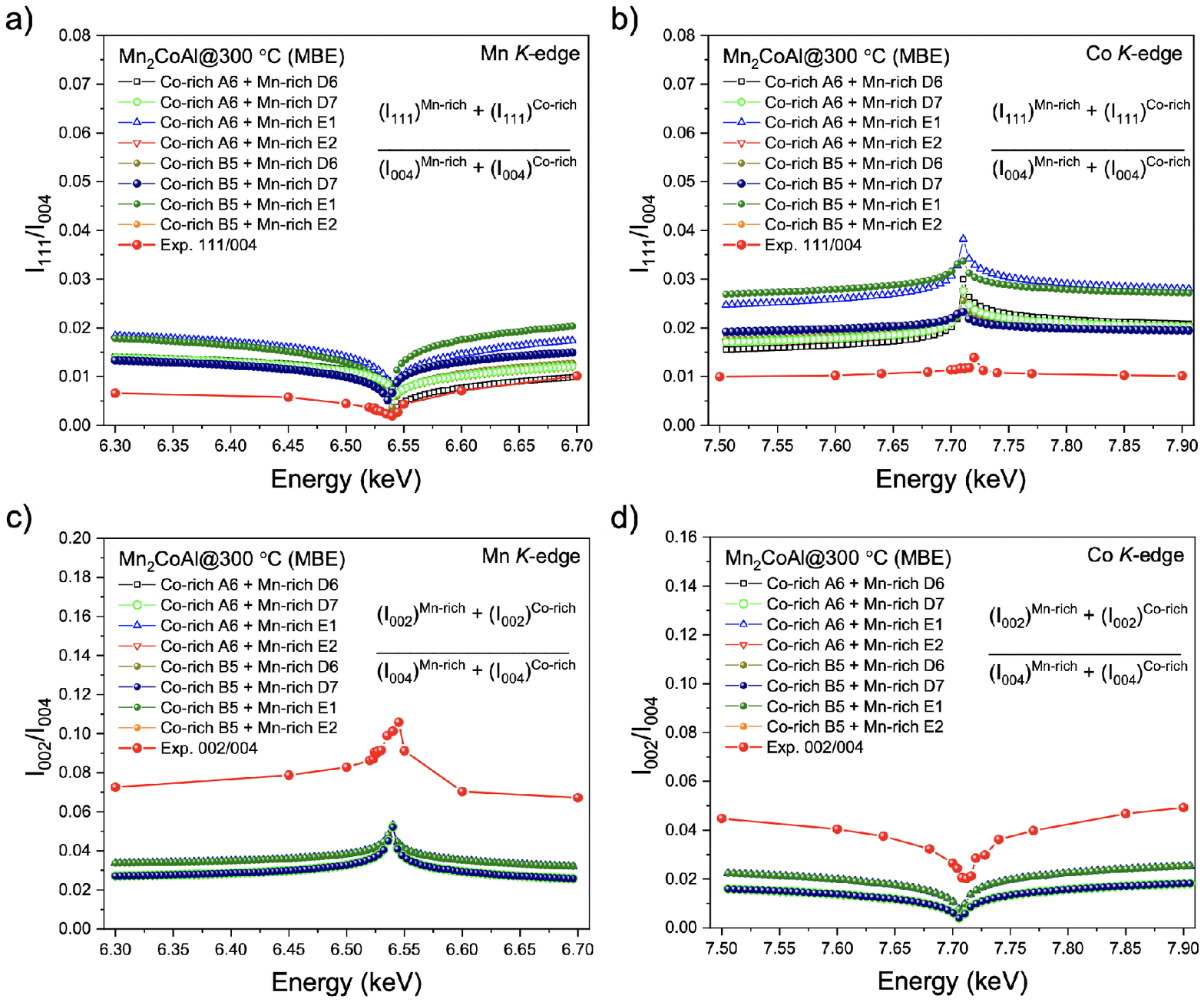}
	\vspace{2.0em}
	\caption{Calculated anomalous XRD normalized by 004 for mixed structure compared with experimentally obtained (a) $I_{111}/I_{004}$ at Mn $K$-edge,  (b) $I_{111}/I_{004}$ at Co $K$-edge, (c) $I_{002}/I_{004}$ at Mn $K$-edge,  and (d)$ I_{002}/I_{004}$ at Co $K$-edge. The combinations were created using selected A-6 and B-5 models in the Co-rich region and another four D-6, D-7, E-1 and E-2 models in the Mn-rich region.}
	\label{FIG:S5}
\end{figure}

Figures~\ref{FIG:S4} and \ref{FIG:S5} show the X-ray energy dependence of diffraction intensities $I_{111}$ and $I_{002}$,
and intensity ratios $I_{111}/I_{004}$ and $I_{002}/I_{004}$,
respectively,
for the structural models in Tables~3 (mixed structures in Co-rich and Mn-rich regions) in the main text
at around both the Co $K$-absorption edge (7.709 keV) and Mn $K$-absorption edge (6.539 keV),
in addition to the experimental data.
%

\subsection{Sample quality}
\begin{figure}
\centering
\includegraphics[width=0.7\linewidth]{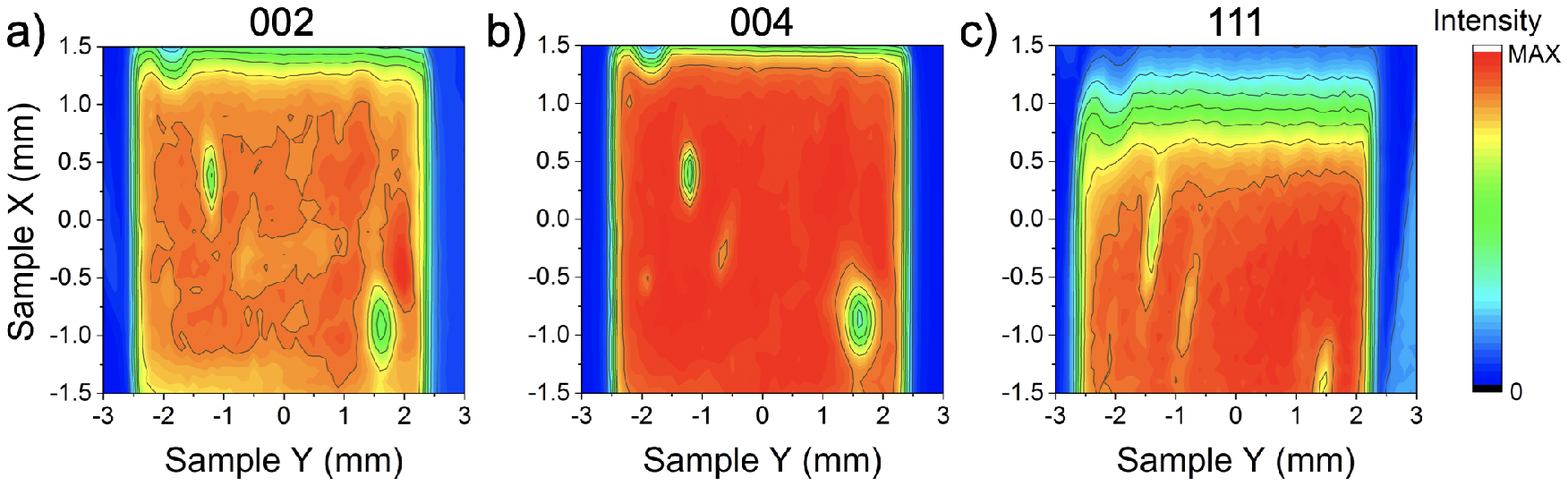}
\vspace{2.0em}
\caption{
Synchrotron XRD color-coded maps of the Mn$_2$CoAl thin film grown by MBE
for the (a) 002, (b) 004, and (c) 111 reflections at the Co $K$-edge.
The peak intensity over a sample region (ca. 3 mm $\times$ 5 mm) was measured to check the homogeneity of the thin film.
}
\label{FIG:AXRDMap}
\end{figure}

Figure~\ref{FIG:AXRDMap} shows color-coded maps of synchrotron XRD of the Mn$_2$CoAl-MBE thin film
at the Co $K$-edge with X-ray beam size of $100 \times 100$ $\mu$m.
For all reflections (002, 004, and 111),
the film showed the almost uniform peak-intensity distribution over the irradiated area
except for a few region the intensities were weak,
which might reflect a film degradation by physical impact during sample handling.
We measured synchrotron anomalous XRD at the center of the sample that showed no film degradation.

\bibliographystyle{elsarticle-num-names}
\bibliography{ActaMater_MCA_AXRD_Supplementary}


\clearpage
\begin{table}[htbp]
	\centering
	\caption{\bf Figures of merit (FOM) for the anomalous XRD of candidate structure models in the Co-rich region. The FOM derived from the summation of all degree of agreements in the first criteria as follows: 1; similar dip or peak shape, 0; no peak or dip shape, -1; opposite profile.}
	\begin{tabular}{ccccccccccccccc}
		\hline
		Exp. peak  &  \multicolumn{7}{c} {Co-rich $XA$} && \multicolumn{6}{c} {Co-rich $L2_1$} \\
		\cline{2-8} \cline{10-15}
		& A-1 & A-2 & A-3 & A-4 & A-5 & A-6 & A-7 && B-1 & \sout{B-2} & B-3 & B-4 & B-5 & \sout{B-6}\\
		\hline
	111@Mn-$K$ & -1 & -1 & -1 & -1 & -1 & 1 & -1& & 1 & 1 & 1 & 1 & 1 & 1 \\
	
	111@Co-$K$ & -1 & -1 & -1 & -1 & -1 & 1 & -1 && 1 & 1 & 1 & 1 & 1 & 1 \\
	
	002@Mn-$K$ & 1 & 1 & 1 & -1 & -1 & 1 & -1 & &-1& 1 & -1 & -1 & 1 & 1 \\
	
	002@Co-$K$ & 1 & 1 & 1 & -1 & -1 & 1 & -1 && -1 & 1 & -1 & -1 & 1 & 1 \\
	
	111/004@Mn-$K$ & -1 & -1 & -1 & -1 & -1 & 1 & -1 && 1 & 1 & 1 & 1 & 1 & 1 \\
	
	111/004@Co-$K$ & -1 & -1 & -1 & -1 & -1 & 1 & -1 && 1 & 1 & 1 & 1 & 1 & 1 \\
	
	002/004@Mn-$K$ & -1 & 1 & -1 & 1 & 1 & 1 & 1 && 1 & 1 & 1 & 1 & 1 & 0 \\
	
	002/004@Co-$K$ & -1 & 1 & -1 & 1 & 1 & 1 & 1 && -1 & 1 & -1 & -1 & 1 & 1 \\
		\hline
		\bf {FOM} & -4 & 0 & -4 & -4 & -4 & \bf {8} & -4 && 2 & \sout{8} & 2 & 2 & \bf {8} & \sout{7}  \\
		\hline
	\end{tabular}
	\label{tab:S3}
\end{table}

\clearpage
\begin{table}[htbp]
	\centering
	\caption{\bf Figures of merit (FOM) for the anomalous XRD of candidate structure models in the Mn-rich region. The FOM derived from the summation of all degree of agreements in the first criteria as follows: 1; similar dip or peak shape, 0; no peak or dip shape, -1; opposite profile.}
	\begin{tabular}{cccccccccccccc}
		\hline
		Exp. peak  &  \multicolumn{10}{c} {Mn-rich $XA$} & &\multicolumn{2}{c} {Mn-rich $L2_1$} \\
		\cline{2-11} \cline{13-14}
		& C-1 & C-2 & C-3 & D-1 & D-2 & D-3 & D-4 & D-5 & D-6 & D-7 & & E-1 & E-2 \\
		\hline
		111@Mn-$K$ & -1 & -1 & 1 & -1 & -1 & -1 & -1 & -1 & 1 & 1 & &  1  &1 \\
		
		111@Co-$K$ & -1 & -1 & -1 & -1 & -1 & -1 & -1 & -1 & 1 & 1 & & 1 & 1 \\
		
		002@Mn-$K$ & -1 & -1 & -1 & 1 & 1 & -1 & 1 & 1 & 1 & 1 & & 1 & 1  \\
		
		002@Co-$K$ & 1 & 1 & -1 & 1 & 1 & 1 & 1 & 1 & 1 & 1 & & 1  & 1 \\
		
		111/004@Mn-$K$ & -1 & -1 & 1 & -1 & -1 &- 1 & -1 & 1 & 1 & 1 & & 1 & 1 \\
		
		111/004@Co-$K$ & 1 & 1 & 1 & 1 & 1 & 1 & 1 & 1 & 1 & 1 & & 1 & 1  \\
		
		002/004@Mn-$K$ & 1 & -1 & 1 & 1 & 1 & -1 & 1 & 1 & 1 & 1 & & 1 & 1  \\
		
		002/004@Co-$K$ & 1 & 1 & 1 & 1 & 1 & -1 & 1 & 1 & 1 & 1 & & 1 & 1  \\
		\hline
		\bf {FOM} & 0 & -2 & 2 & 2 & 2 & -4 & 2 & 4 & \bf {8} & \bf {8} & & \bf {8} & \bf {8}  \\
		\hline
	\end{tabular}
	\label{tab:S4}
\end{table}

\clearpage

\clearpage

\clearpage

\clearpage

\clearpage

\clearpage
